\documentclass{article}
\usepackage{spconf,amsmath,graphicx,here}
\usepackage{url,setspace,booktabs}
\usepackage{amssymb}
\usepackage[breaklinks=true]{hyperref}
\usepackage[anythingbreaks]{breakurl}

\usepackage[
backend=biber,
style=ieee,
doi=false,isbn=false,url=false,eprint=false
]{biblatex} 

\defbibheading{bibliography}[\refname]{}

\DeclareSourcemap{
	\maps[datatype=bibtex, overwrite=true]{
		\map{
			\step[fieldsource=booktitle,
			match=\regexp{.*Interspeech.*},
			replace={Proc. Interspeech}]
			\step[fieldsource=journal,
			match=\regexp{.*INTERSPEECH.*},
			replace={Proc. Interspeech}]
			\step[fieldsource=booktitle,
			match=\regexp{.*ICASSP.*},
			replace={Proc. ICASSP}]
			\step[fieldsource=booktitle,
			match=\regexp{.*icassp_inpress.*},
			replace={ICASSP (in press)}]
			\step[fieldsource=booktitle,
			match=\regexp{.*International.*Conference.*on.*Acoustics,.*Speech.*and.*Signal.*Processing.*},
			replace={Proc. ICASSP}]
			\step[fieldsource=booktitle,
			match=\regexp{.*International.*Conference.*on.*Learning.*Representations.*},
			replace={Proc. ICLR}]
			\step[fieldsource=booktitle,
			match=\regexp{.*International.*Conference.*on.*Machine.*Learning.*},
			replace={Proc. ICML}]
			\step[fieldsource=booktitle,
			match=\regexp{.*Automatic.*Speech.*Recognition.*and.*Understanding.*},
			replace={Proc. ASRU}]
			\step[fieldsource=booktitle,
			match=\regexp{.*Spoken.*Language.*Technology.*},
			replace={Proc. SLT}]
			\step[fieldsource=booktitle,
			match=\regexp{.*Speech.*Synthesis.*Workshop.*},
			replace={Proc. SSW}]
			\step[fieldsource=booktitle,
			match=\regexp{.*workshop.*on.*speech.*synthesis.*},
			replace={Proc. SSW}]
			\step[fieldsource=booktitle,
			match=\regexp{.*Advances.*in.*neural.*information.*processing.*},
			replace={Proc. NeurIPS }]
			\step[fieldsource=booktitle,
			match=\regexp{.*Workshop.*on.* Applications.* of.* Signal.*Processing.*to.*Audio.*and.*Acoustics.*},
			replace={Proc. WASPAA}]
			\step[fieldsource=publisher,
			match=\regexp{.+},
			replace={{}}]
			\step[fieldsource=month,
			match=\regexp{.+},
			replace={{}}]
			\step[fieldsource=location,
			match=\regexp{.+},
			replace={{}}]
			\step[fieldsource=address,
			match=\regexp{.+},
			replace={{}}]
			\step[fieldsource=organization,
			match=\regexp{.+},
			replace={{}}]
		}
	}
}

\title{Non-autoregressive sequence-to-sequence voice conversion}
\name{Tomoki Hayashi$^{1,2}$, Wen-Chin Huang$^2$, Kazuhiro Kobayashi$^{1,2}$, Tomoki Toda$^2$}
\address{
$^{1}$TARVO Inc., Japan, $^{2}$Nagoya University, Japan, \\
 \url{hayashi.tomoki@g.sp.m.is.nagoya-u.ac.jp}
}

\begin{document}
\ninept
\maketitle
\begin{abstract}
This paper proposes a novel voice conversion (VC) method based on non-autoregressive sequence-to-sequence (NAR-S2S) models.
Inspired by the great success of NAR-S2S models such as FastSpeech in text-to-speech (TTS), we extend the FastSpeech2 model for the VC problem.  
We introduce the convolution-augmented Transformer (Conformer) instead of the Transformer, making it possible to capture both local and global context information from the input sequence.
Furthermore, we extend variance predictors to variance converters to explicitly convert the source speaker's prosody components such as pitch and energy into the target speaker. 
The experimental evaluation with the Japanese speaker dataset, which consists of male and female speakers of 1,000 utterances, demonstrates that the proposed model enables us to perform more stable, faster, and better conversion than autoregressive S2S (AR-S2S) models such as Tacotron2 and Transformer.
\end{abstract}
\section{Introduction}
\label{sec:intro}
Voice conversion (VC) is a technique to make it possible
to convert speech of the source speaker into that of the target speaker while keeping the linguistic content~\cite{toda2014augmented}.
VC is an essential technique for various applications such as a speaking aid system~\cite{kain2007improving}, a vocal effector for singing voice \cite{villavicencio2010applying,kobayashi2014voice}, and a real-time voice changer for the online communication~\cite{toda2012implementation}.
Conventional VC methods~\cite{toda2007voice,kobayashi2014statistical,sun2015voice} are based on the analysis-synthesis framework \cite{kawahara2006straight,morise2016world}, where speech is decomposed into a spectral feature, pitch, and aperiodicity components.
Since the conventional methods convert each component frame-by-frame, it is difficult to convert the duration or prosody of speech, resulting in limited conversion quality.

% 最近のSeq2Seqがすごい
With the improvement of deep learning techniques, attention-based sequence-to-sequence (S2S) VC models have attracted attention~\cite{tanaka2019atts2s,kameoka2018convs2s,zhang2019improving,huang2019voice,kameoka2020many}.
Thanks to the attention mechanism~\cite{bahdanau2014neural}, S2S-based VC models can learn the alignment between the source sequence and the target sequence in a data-driven manner, making it possible to convert the duration and prosody components.
However, the attention-based S2S models require relatively a large amount of the training data to get a good alignment~\cite{huang2019voice}, causing conversion errors such as the deletion and repetition~\cite{kameoka2018convs2s}.
To address this issue, various approaches have been proposed, such as the attention regularization training~\cite{tanaka2019atts2s}, monotonic attention mechanism~\cite{kameoka2020many}, and the use of the automatic speech recognition (ASR) or text-to-speech (TTS) pretraining model~\cite{huang2019voice}.

Recently, non-autoregressive S2S (NAR-S2S) end-to-end TTS (E2E-TTS) models have achieve the great performances~\cite{peng2019parallel,ren2019fastspeech,ren2020fastspeech,lancucki2020fastpitch}.
The NAR-S2S models enable us to generate the output sequence much faster than autoregressive S2S (AR-S2S) and alleviate generation errors derived from the attention mechanism.
Furthermore, FastSpeech2~\cite{ren2020fastspeech} and FastPitch~\cite{lancucki2020fastpitch} have achieved faster, better, and more stable generation than AR-S2S models by using extra acoustic features such as pitch and energy.

%% 本論文の貢献
This paper proposes a novel voice conversion method based on NAR S2S model for the one-to-one VC.
Inspired by the great successes of NAR-E2E-TTS models~\cite{ren2019fastspeech,ren2020fastspeech,lancucki2020fastpitch}, we extend FastSpeech2~\cite{ren2020fastspeech} for the VC problem.
The contributions are summarized as follows:
\begin{itemize}
    \item We introduce the convolution-augmented Transformer (Conformer)~\cite{gulati2020conformer} architecture instead of Transformer in FastSpeech2~\cite{ren2020fastspeech}. The Conformer architecture enables us to capture both local and global context information from the input sequence, making the conversion quality better.
    \item We extend variance predictors, which predict pitch and energy from the token embedding, into variance converters, converting the source speaker's pitch and energy into the target speaker's one. The variance converters enable us to convert the prosody components more accurately.
    \item We perform an experimental evaluation using a Japanese speaker dataset, which consists of two speakers of 1000 utterances. The experimental results demonstrate that the proposed method can outperform the conventional AR-S2S models in both objective and subjective evaluation.
\end{itemize}

\section{Proposed method}
\subsection{Overview}
\begin{figure}[t]
  \centering
  \centerline{\includegraphics[width=1\columnwidth]{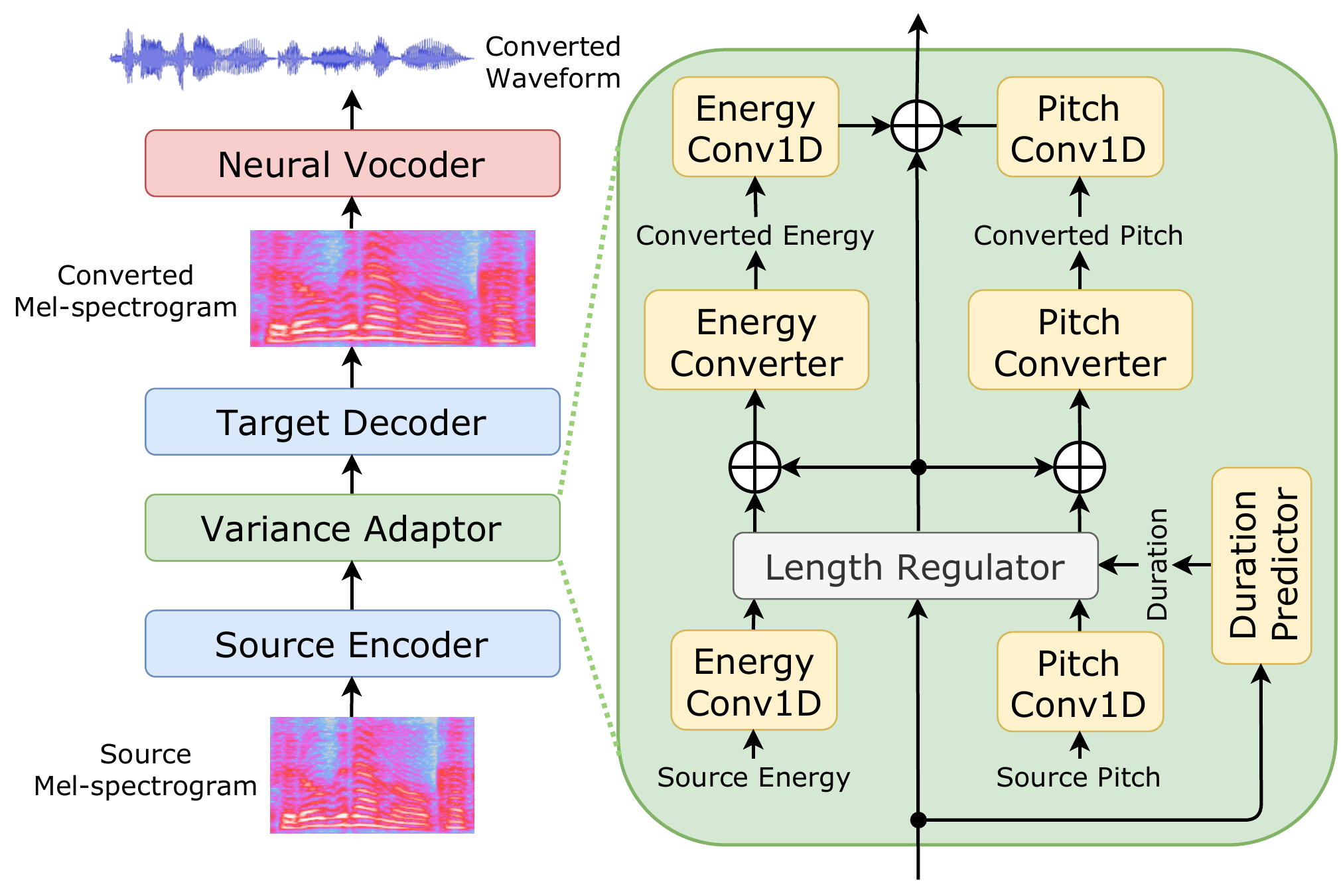}}
  \vspace{-1mm}
  \caption{\it An overview of the proposed method.}
  \label{fig:overview}
\end{figure}

% Overview
The overview of the proposed method is shown in Fig.~\ref{fig:overview}.
% --------------------------------------------- %
% --------------------------------------------- %
The proposed method consists of a source encoder, a duration predictor, a length regulator, variance converters, a target decoder, and a neural vocoder.
The source encoder consists of Conformer-blocks, and it converts Mel-spectrogram of the source speaker into a hidden representation sequence.
The input Mel-spectrogram of the source speaker is normalized to be mean zero and variance one over the source speaker's training data.
To reduce the length of the Mel-spectrogram, we introduce a reduction factor scheme~\cite{wang2017tacotron}, which concatenates the several frames into one frame.
The duration predictor predicts each frame's duration from the encoder hidden representation (See section~\ref{ssec:duration_predictor}).
To solve the mismatch between the length of the source and the target, the length regulator replicates each frame of the source components with duration information~\cite{ren2019fastspeech}.
The variance converters convert the source speaker's pitch and energy into that of the target speaker (See section~\ref{ssec:variance_converter}).
Then, the target decoder, which consists of Conformer blocks and Postnet~\cite{shen2017tacotron2}, predicts the Mel-spectrogram of the target speaker with replicated encoder hidden representations, converted pitch, and converted energy, and then the Postnet~\cite{shen2017tacotron2} refines it.
We use the ground-truth of duration, pitch, and energy as the decoder inputs during the training.
The encoder and decoder are optimized to minimize the mean absolute error between the predicted Mel-spectrogram and the target Mel-spectrogram.
We use the same reduction factor scheme and normalization for the target Mel-spectrogram, but the normalization is performed over the target speaker's training data.
Finally, the neural vocoder generates the waveform with the converted Mel-spectrogram (See section~\ref{ssec:neural_vocoder}).

\subsection{Conformer-block}
\begin{figure}[t]
  \centering
  \centerline{\includegraphics[width=1\columnwidth]{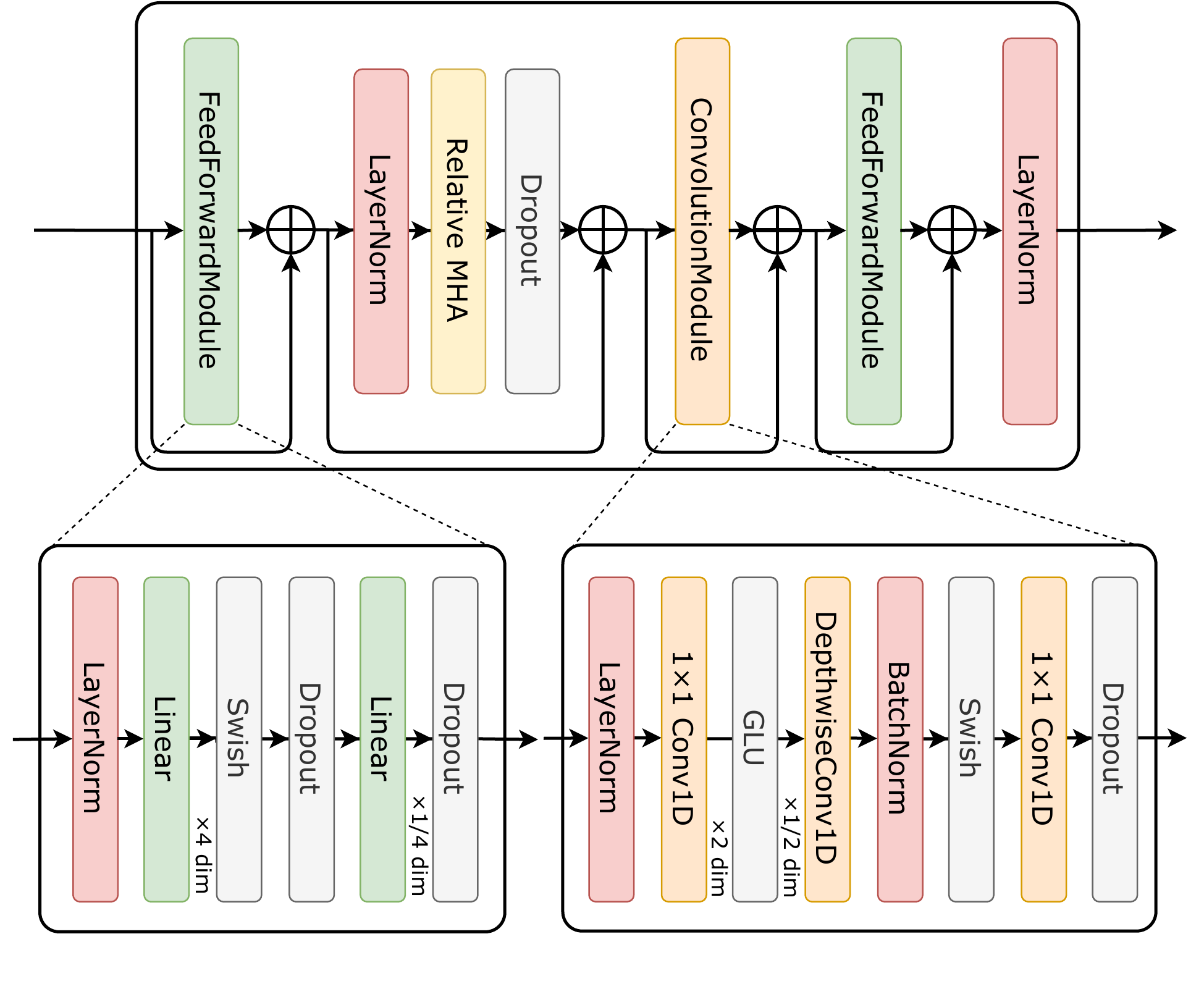}}
  \vspace{-1mm}
  \caption{\it An architecture of the Conformer-block.}
  \label{fig:conformer}
\end{figure}

The Conformer is a convolution-augmented Transformer, which achieves state-of-the-art performance in the field of ASR~\cite{gulati2020conformer}.
The architecture of the Conformer block is illustrated in Fig.~\ref{fig:conformer}.
The Conformer block basically consists of feed-forward modules, a multi-head attention layer, and a convolution module.
The feed-forward module consists of a LayerNorm layer, a linear layer with Swish activation~\cite{ramachandran2017searching} and dropout, and a linear layer with dropout.
The first linear layer expands the dimension four times, and the second one projects back to the original input dimension.
After both linear layers, we multiply the outputs by $0.5$ following the method described in the original Conformer paper~\cite{gulati2020conformer}.
The multi-headed attention uses the relative sinusoidal positional encoding introduced in Transformer-XL~\cite{dai2019transformer}, which makes the model more robust to the various inputs of different lengths.
The convolution module consists of a LayerNorm layer, a $1\times1$ 1D convolution layer with a gated linear unit (GLU) activation~\cite{dauphin2017language}, and a 1D depth-wise convolution layer.
The depth-wise convolution layer is followed by a 1D batch normalization layer, a Swish activation, and a $1\times1$ 1D convolution layer with dropout.

% --------------------------------------------- %
% --------------------------------------------- %

\subsection{Duration predictor}
\label{ssec:duration_predictor}
We follow the variance predictor architecture in \cite{ren2020fastspeech}, which consists of several layers, including 1D convolution and ReLU activation followed by LayerNorm and dropout. 
The duration predictor is trained to minimize the mean square error between the predicted duration and the ground-truth in the log domain.
To obtain the ground-truth, we train the AR-Transformer model~\cite{huang2019voice} and then calculate the duration from the attention weights of the source-target diagonal attention by following the manner in FastSpeech~\cite{ren2019fastspeech}.
We use teacher forcing in calculating attention weights to obtain the duration, the length of which is matched with the target Mel-spectrogram.
We can use any attention-based S2S model to obtain the ground-truth duration, but in our preliminary experiments, Transformer gives us more concrete alignment than the Tacotron2-based model.

% % --------------------------------------------- %
% \begin{figure}[t]
%   \centering
%   \centerline{\includegraphics[width=1\columnwidth]{dummy.eps}}
%   \vspace{0mm}
%   \caption{\it An architecture of the duration predictor.}
%   \label{fig:variance_predictor}
% \end{figure}
% % --------------------------------------------- %

% --------------------------------------------------------- %
\begin{table*}[t!]
\begin{center}
\vspace{0mm}
\vspace{0mm}
\caption{\it Objective evaluation results. The values in MCD and log $F_0$ RMSE represent the mean and the standard deviation.}
\begin{tabular}{lcccccc}
& \multicolumn{3}{c}{ Male $\longrightarrow$ Female } & \multicolumn{3}{c}{ Female $\longrightarrow$ Male } \\
\toprule
Method & MCD [dB] & log $F_0$ RMSE & CER [\%] & MCD [dB] & log $F_0$ RMSE & CER [\%] \\
\midrule
Ground-truth & N/A & N/A & 10.6 & N/A & N/A & 11.3 \\
\midrule
Tacotron2 & 6.33 $\pm$\ 0.36 & 0.16 $\pm$\ 0.03 & 24.9 & 5.30 $\pm$\ 0.20 & 0.22 $\pm$\ 0.04 & 19.4 \\
Transformer & 6.11 $\pm$\ 0.63 & 0.16 $\pm$\ 0.03 & 19.6 & 5.28 $\pm$\ 0.79 & 0.21 $\pm$\ 0.05 & 21.1 \\
Proposed & \textbf{5.71} $\pm$\ \textbf{0.32} & \textbf{0.15} $\pm$\ \textbf{0.02} & \textbf{15.3} & \textbf{4.90} $\pm$\ \textbf{0.22} & \textbf{0.21} $\pm$\ \textbf{0.04} & \textbf{14.8} \\
\bottomrule
\end{tabular}

\label{tb:obj_result}
\end{center}
\end{table*}
% --------------------------------------------------------- %

\subsection{Variance converter}
\label{ssec:variance_converter}
% Variance converter explanation
The variance converters follow the same architecture of the duration predictor except for the inputs.
We extract continuous log pitch and energy from the waveform of both the source speaker and the target speaker and then normalize them to be mean zero and variance one over each speaker's training data. 
We use Dio and Stonemask algorithm~\cite{morise2010rapid} with PyWorld\footnote{\url{https://github.com/JeremyCCHsu/Python-Wrapper-for-World-Vocoder}} to extract pitch.
The extracted pitch or energy is inputted into an additional 1D convolution layer to become the same dimension of the encoder hidden representations, and we use their summation as the inputs of the variance converters.
Unlike FastSpeech2~\cite{ren2020fastspeech}, we do not perform discretization in the variance converters since we can use the source pitch and energy as the input.
We optimize the variance converters to minimize the mean square errors and stop the gradient flow from the pitch converter to the encoder to avoid overfitting. 

\subsection{Neural vocoder}
\label{ssec:neural_vocoder}
The proposed model can use any neural vocoder which accepts 
Mel-spectrogram as the inputs.
In this study, we use Parallel WaveGAN~\cite{yamamoto2020parallel} as the neural vocoder since it can generate a waveform faster than the real-time and can be trained with a limited amount of the training data (e.g., less than one hour).
We use an open-source implementation available in Github~\cite{hayashi_pwg}.

\section{Experimental Evaluation}

\subsection{Experimental condition}
To demonstrate the proposed method's performance, we conducted experimental evaluations using the Japanese speech parallel dataset.
The dataset consists of one female speaker and one male speaker of 1,000 utterances.
The dataset is recorded in a low reverberation and quiet room, and the sampling rate is a 24 kHz sampling rate.
We used 950 utterances for the training, 25 utterances for the validation, and 25 utterances for the evaluation.

The implementation was based on the open-source ESPnet toolkit~\cite{hayashi2020espnet}.
We extracted an 80-dimensional Mel-spectrogram with 2,048 FFT points, 300 shift points, and a hanning window of 1,200 points. 
The reduction factor was set to three for both the source and the target.
The number of encoder blocks and decoder blocks was set to four, and heads in multi-headed attention were set to two.
The attention dimension was set to 384, and the kernel size of the convolutional module in the Conformer block is seven.
The number of channels in the duration predictor and variance converters was set to 256.
The number of layers of the duration predictor and the energy converter was two, and their kernel size was three.
The number of layers of the pitch converter was five, and the kernel size in the pitch converter was five.
We trained 100k iterations using Noam optimizer~\cite{vaswani2017attention}, and the warmup steps were set to 4,000.
The neural vocoder was trained with the ground-truth Mel-spectrogram of the target speaker based on \url{parallel_wavegan.v1} configuration available in \cite{hayashi_pwg}.
As baseline methods, we used the following models:
\begin{description}
    \item \textbf{Tacotron2}: Tacotron2-based attention S2S VC. We followed the setting described in \cite{tanaka2019atts2s} with the guided attention loss and the source and target reconstruction loss. The Mel-spectrogram was used instead of the WORLD~\cite{morise2016world}-based acoustic features.   
    \item \textbf{Transformer}: Transformer-based attention S2S VC~\cite{huang2019voice}. We followed the original implementation in ESPnet, but we added the LayerNorm layer in the front of each Transformer-block, which makes the training more stable.
\end{description}
We use the same feature extraction setting, reduction factors, and the neural vocoder for the baselines.
The evaluation samples are publicity available in \cite{hayashi_sample}.

\subsection{Objective evaluation results}
As the objective evaluation metrics, we used Mel-cepstrum distortion (MCD), root mean square error (RMSE) of log $F_0$, and character error rate (CER).
To get the alignment between the prediction and the reference, we used dynamic time warping (DTW).
To calculate the MCD, we calculated 0-34 order Mel-cepstrum with SPTK toolkit\footnote{\url{http://sp-tk.sourceforge.net/}}.
The CER was calculated by the Transformer-based ASR model trained on the corpus of spontaneous Japanese (CSJ)~\cite{maekawa2003corpus}, which was provided by ESPnet~\cite{watanabe2018espnet}.

\begin{figure}[t!]
  \centering
  \centerline{\includegraphics[width=1\columnwidth]{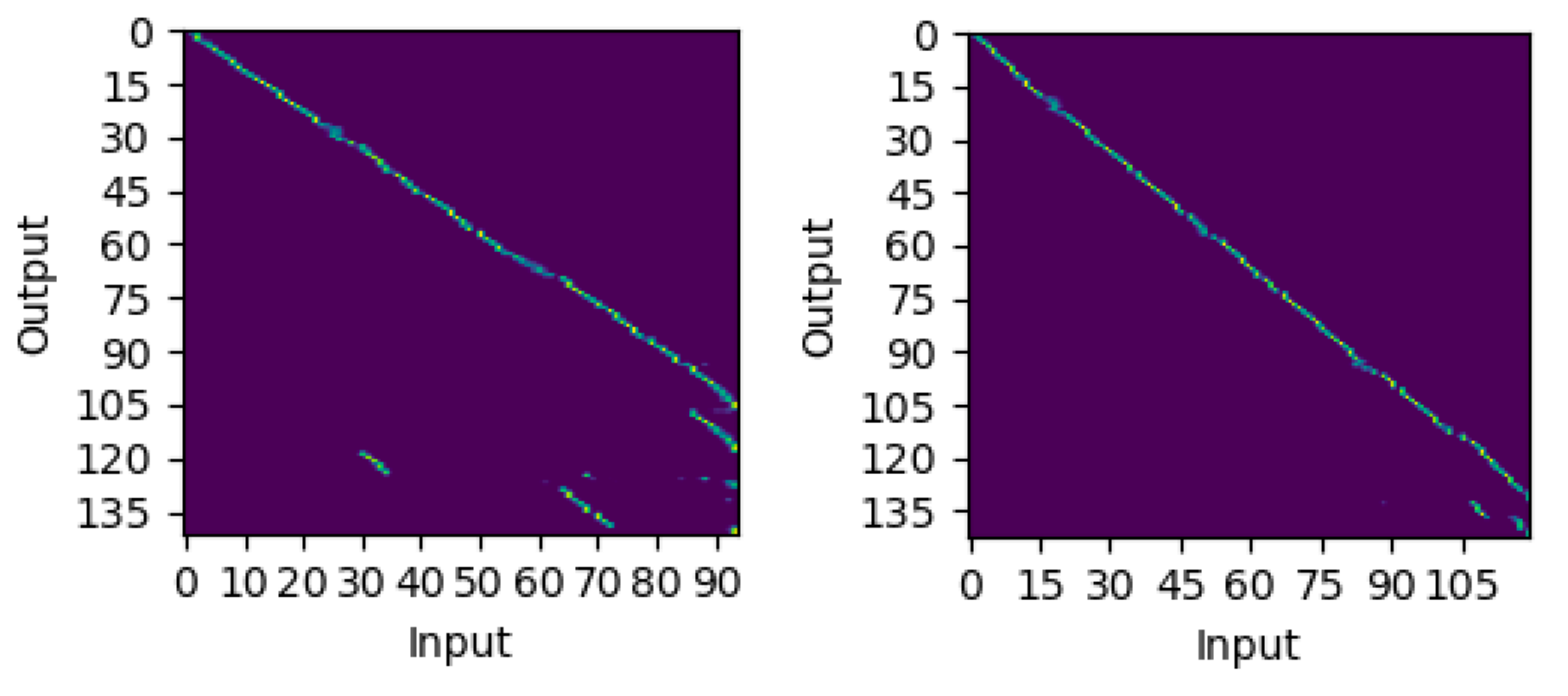}}
  \vspace{-1mm}
  \caption{\it Examples of alignment failures in Transformer.}
  \label{fig:align_failure}
\end{figure}

The objective evaluation result is shown in Table~\ref{tb:obj_result}.
The results show that the proposed method outperformed the baselines in both male to female and female to male conversions.
The Transformer model usually gave better results than the Tacotron2 model, but sometimes the decoding was failed due to the alignment error, as shown in Fig.~\ref{fig:align_failure}
This caused the repetition of speech, and it resulted in a higher standard deviation of MCD or worse CER.
On the other hand, the proposed method could alleviate the above problem thanks to the NAR architecture.
% --------------------------------------------- %
% --------------------------------------------- %

Table~\ref{tb:speed_comp} shows the conversion speed comparison.
The evaluation was conducted with eight threads of CPUs (Xeon Gold, 3.00 GHz) and a single GPU (NVIDIA TITAN V).
% --------------------------------------------------------- %
\begin{table}[t!]
\begin{center}
\vspace{0mm}
\vspace{0mm}
\caption{\it Conversion speed comparison. The values represents the mean of the standard deviation. The time of waveform generation was not included.}
\vspace{1mm}
\begin{tabular}{lcc}
& \multicolumn{2}{c}{Conversion speed [frames / sec]} \\
\toprule
Method & On CPU & On GPU \\
\midrule
Tacotron2 & 1240.0 $\pm$\ 47.4 & 2147.3 $\pm$\ 200.6 \\
Transformer & 225.6 $\pm$\ 13.2 & 359.2 $\pm$\ 18.0 \\
Proposed & \textbf{3640.1} $\pm$\ \textbf{745.6} & \textbf{16723.7} $\pm$\ \textbf{3176.7} \\
\bottomrule
\end{tabular}
\label{tb:speed_comp}
\end{center}
\end{table}
% --------------------------------------------------------- %
The result shows that the proposed method can perform the conversion mush faster than AR S2S models, thanks to the NAR architecture.
To realize the real-time streaming voice conversion, we will consider the extension to the streaming model in future work.

% --------------------------------------------------------- %
\begin{table*}[t!]
\begin{center}
\vspace{0mm}
\vspace{0mm}
\caption{\it Subjective evaluation results. The values represent the mean and 95 \% confidence interval.}
\begin{tabular}{lcccccc}
& \multicolumn{2}{c}{ Male $\longrightarrow$ Female } & \multicolumn{2}{c}{ Female $\longrightarrow$ Male } & \multicolumn{2}{c}{ Average } \\
\toprule
Method & Naturalness & Similarity & Naturalness & Similarity & Naturalness & Similarity \\
\midrule
Groundtruth & 3.67 $\pm$\ 0.14 & N/A & 4.51 $\pm$\ 0.14 & N/A & 4.12 $\pm$\ 0.09 & N/A \\
Tacotron2 & 2.41 $\pm$\ 0.13 & 58\% $\pm$\ 7\% & 3.01 $\pm$\ 0.15 & 52\% $\pm$\ 7\% & 2.71 $\pm$\ 0.10 & 55\% $\pm$\ 5\% \\
Transformer & 3.09 $\pm$\ 0.14 & \textbf{85\%} $\pm$\ \textbf{5\%} & 3.43 $\pm$\ 0.15 & 76\% $\pm$\ 6\% & 3.27 $\pm$\ 0.11 & 81\% $\pm$\ 4\% \\
Proposed & \textbf{3.26} $\pm$\ \textbf{0.14} & \textbf{85\%} $\pm$\ \textbf{5\%} & \textbf{3.64} $\pm$\ \textbf{0.15} & \textbf{78\%} $\pm$\ \textbf{6\%} & \textbf{3.47} $\pm$\ \textbf{0.10} & \textbf{82\%} $\pm$\ \textbf{4\%} \\
\bottomrule
\end{tabular}
\label{tb:sub_result}
\end{center}
\end{table*}
% --------------------------------------------------------- %

\subsection{Subjective evaluation results}
We conducted subjective evaluation tests on naturalness and speaker similarity to check the perceptual quality.
For naturalness, each subject evaluated 50 samples and rated the naturalness of each sample on a 5-point scale: 5 for excellent, 4 for good, 3 for fair, 2 for poor, and 1 for bad.
For speaker similarity, each subject evaluated 40 pairs of the target sample and the converted sample to judge whether the presented samples were produced by the same speaker with confidence (e.g., sure or not sure).
The number of subjects was 36 of Japanese native speakers.

Table~\ref{tb:sub_result} shows the subjective evaluation results.
The result demonstrated that the proposed method outperformed the conventional AR-S2S models in both conversion conditions, showing the proposed method's effectiveness.

\subsection{Ablation study}
Next, we conducted an ablation study to check the effectiveness of each extension.
The results of the objective evaluation is shown in Table~\ref{tb:ablation_result}.
% --------------------------------------------------------- %
\begin{table}[t!]
\begin{center}
\vspace{0mm}
\vspace{0mm}
\caption{\it Ablation study results on the male to female conversion. Conf. and V.C. represents the use of Conformer and variance converters. We use the Transformer-block instead of the Conformer-block when no Conformer was used.}
\scalebox{0.93}{\renewcommand\arraystretch{1}{%
\begin{tabular}{cccccc}
\toprule
Conf. & V.C. & Teacher & MCD & log $F_0$ RMSE & CER \\
\midrule
\checkmark & \checkmark & Transformer & \textbf{5.71} $\pm$\ \textbf{0.32} & \textbf{0.15} $\pm$\ \textbf{0.02} & \textbf{15.3} \\
& \checkmark & Transformer & 5.82 $\pm$\ 0.33 & 0.15 $\pm$\ 0.02 & 16.1 \\
& & Transformer & 6.08 $\pm$\ 0.63 & 0.16 $\pm$\ 0.02 & 20.4 \\
\checkmark & \checkmark & Tacotron2 & 5.84 $\pm$\ 0.29 & 0.15 $\pm$\ 0.02 & 19.7 \\
& \checkmark & Tacotron2 & 5.87 $\pm$\ 0.33 & 0.15 $\pm$\ 0.02 & 17.8 \\
\bottomrule
\end{tabular}
}}
\label{tb:ablation_result}
\end{center}
\end{table}
% --------------------------------------------------------- %
% --------------------------------------------- %
\begin{figure}[t!]
  \centering
  \centerline{\includegraphics[width=1\columnwidth]{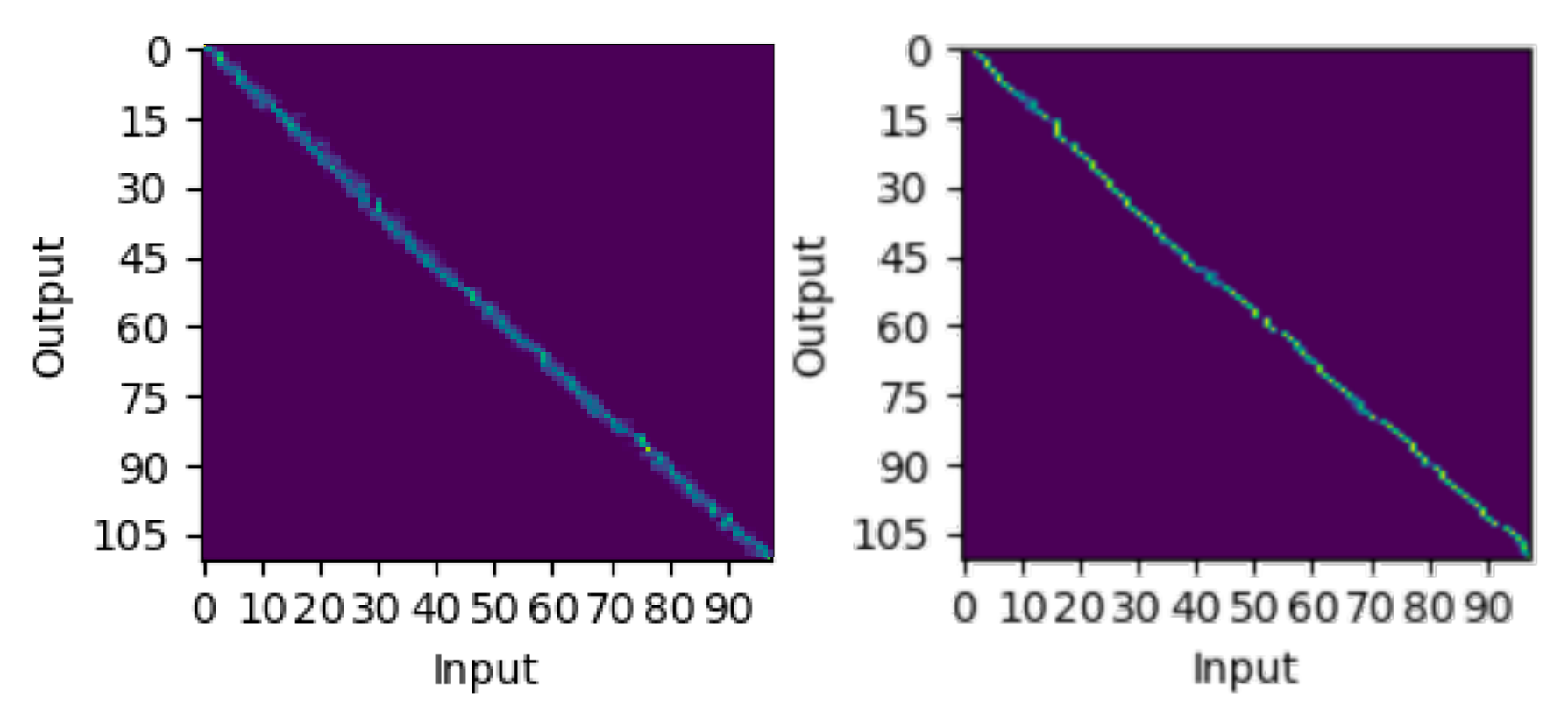}}
  \vspace{-1mm}
  \caption{\it Comparison of alignments Tacotron2 (left, the focus rate \cite{ren2019fastspeech} is 0.41) vs. Transformer (right, the focus rate is 0.93). Transformer's alignment was obtained from the most diagonal source-target attention.}
  \label{fig:align_comp}
\end{figure}
% --------------------------------------------- %
The result showed that both the use of Conformer and that of variance converters could improve the performance.
If we do not use the Conformer and variance converters (this case is similar to FastSpeech architecture), the performance is almost the same as the Transformer.
This implied that the NAR model's upper bound without variance converters are the teacher model.
The result also showed that the teacher model's duration affected the performance.
One of the reasons is that the Transformer can generate more concrete alignment between the output and the input, i.e., accurate duration than the Tacotron2. 
The comparison of the attention weights is shown in Fig.\ref{fig:align_comp}.
Since NAR S2S models do not need to learn data-driven alignments, it is expected that the required training data is much smaller than AR S2S models.
To take the above advantage, we will consider another method to obtain the duration in future work.

% \subsection{Controllability analysis}
% なんか乗ると良い

\section{Conclusion}
In this paper, we proposed a novel NAR S2S VC method based on FastSpeech2.
The experimental evaluation results showed that the proposed method outperformed the conventional AR-S2S models and converted speech much faster than them.
In future work, we will consider the extension to the streaming model with a causal architecture, the obtainment of the ground-truth duration without AR-S2S models, and extension to many-to-many voice conversion with learnable embedding.

\section{Acknowledgement}
This work was partly supported by JSPS KAKENHI Grant-in-Aid for JSPS Research Fellow Number 19K20295, and by JST, CREST Grant Number JPMJCR19A3.

% \bibliographystyle{IEEEbib}
% \bibliography{strings,refs}

\section{References}
{
% \setstretch{0.82}
\printbibliography
}

\end{document}